\documentclass{article}
\usepackage{jheppub}
 \normalsize
\def\lsim{\raise0.3ex\hbox{$\;<$\kern-0.75em\raise-1.1ex\hbox{$\sim\;$}}}
\def\gsim{\raise0.3ex\hbox{$\;>$\kern-0.75em\raise-1.1ex\hbox{$\sim\;$}}}

\usepackage{graphics}
\def\met{\slashed E_T}
\usepackage{epsfig}
\usepackage{slashed}
\usepackage[utf8]{inputenc}
\usepackage{multirow}
\usepackage{pstricks}
\usepackage{dcolumn}
\newcommand{\be}{\begin{eqnarray}}
\newcommand{\ee}{\end{eqnarray}}

\def\bea{\begin{eqnarray}}
\def\eea{\end{eqnarray}}
\usepackage{epsfig}
\usepackage{array}
\usepackage{booktabs}
\usepackage{slashed}

\begin{document}
\title{Search for Mono-Higgs Signals  at the LHC\\
 in the $B-L$ Supersymmetric Standard Model }
\author{W. Abdallah$^{1,2}$, A. Hammad$^{1}$, S. Khalil$^{1}$ and S. Moretti$^{3}$ }
\vspace*{0.2cm}
\affiliation{$^1$Center for Fundamental Physics, Zewail City of Science and Technology, 6 October City, Giza 12588, Egypt.\\
$^2$Department of Mathematics, Faculty of Science, Cairo
University, Giza 12613, Egypt.\\
$^3$School of Physics and Astronomy, University of Southampton,
Highfield, Southampton SO17 1BJ, UK.
}
\emailAdd{wabdallah@zewailcity.edu.eg}
\emailAdd{ahammad@zewailcity.edu.eg}
\emailAdd{skhalil@zewailcity.edu.eg}
\emailAdd{s.moretti@soton.ac.uk}
\date{\today}

\abstract{
We study mono-Higgs signatures emerging in the $B-L$ supersymmetric standard model induced by new channels
not present in the minimal supersymmetric standard model, i.e., via  topologies in which the mediator
is either a heavy $Z'$, with mass of ${\cal O}(2~{\rm TeV})$, or an intermediate $h'$ (the lightest CP-even Higgs state of $B-L$ origin), with mass of ${\cal O}(0.2~{\rm TeV})$. The mono-Higgs probe considered is the SM-like Higgs state recently discovered at 
the large hadron collider, so as to enforce its mass reconstruction for background reduction purposes. With this in mind, its two cleanest signatures are selected: $\gamma\gamma$ and $ZZ^*\to 4l$ ($l=e,~\mu$). We show how both of
these can be accessed with foreseen energy and luminosity options using a dedicated kinematic analysis performed in presence of partonic, showering, hadronisation and detector effects.}
\maketitle
\section{Introduction}
\label{sec:intro}

The increased pressure exercised by current experimental data on the parameter space of the Minimal Supersymmetric Standard Model (MSSM) combined with the unsatisfactory theoretical situation highlighting  a  severe fine-tuning problem therein (also known as the small hierarchy problem)
calls 
for the phenomenological exploration of non-minimal constructs of Supersymmetry (SUSY) better
compatible with current data than the MSSM yet similarly predictive and appealing theoretically. Because of the 
well established existence of non-zero neutrino masses, a well motivated path to follow in this direction is to consider the $B-L$ Supersymmetric Standard Model
(BLSSM). Herein, (heavy) right-handed neutrino
superfields are introduced in order to implement a type I
seesaw mechanism, which provides an elegant solution for the existence and
smallness of the (light) left-handed neutrino masses. Right-handed neutrinos can 
naturally be implemented in the BLSSM, which is based on the gauge group $SU(3)_C \times
SU(2)_L \times U(1)_Y \times U(1)_{B-L}$, hence the
simplest generalisation of the SM gauge group (through an additional $U(1)_{B-L}$ symmetry). In this model, it has been shown that the scale of $B-L$ symmetry breaking is related to the SUSY breaking
scale \cite{Khalil:2007dr}, so that this SUSY realisation predicts several testable signals at the Large Hadron Collider (LHC),
not only in the sparticle domain but also in the $Z'$  (a $Z'$ boson in fact emerges from the $U(1)_{B-L}$
breaking), Higgs (an additional singlet state is economically introduced here, breaking the $U(1)_{B-L}$ group) and (s)neutrino sectors \cite{Khalil:2006yi,B-L-LHC,PublicPapers}. Furthermore, other than assuring its testability
at the LHC, in fact, in a richer form than the MSSM (because of the additional (s)particle states), the BLSSM also alleviates the aforementioned
little hierarchy problem
of the MSSM, as both the additional singlet Higgs state and right-handed (s)neutrinos
\cite{BLMSSM-Higgs,O'Leary:2011yq,Basso:2012ew,Elsayed:2012ec,Khalil:2015naa}
release additional parameter space from the LEP, Tevatron and LHC constraints.
Finally, interesting results on the ability of the BLSSM to emulate the Higgs boson signals isolated at the LHC Run 1
have also emerged, including the possibility of explaining possible anomalies hinting 
at a second Higgs peak in the ATLAS and CMS data samples \cite{Abdallah:2014fra}.
A Dark Matter (DM) candidate within the BLSSM which is plausibly different from the MSSM one exists as well \cite{Basso:2012gz}. 

The best probe of a DM signal at the LHC is via the mono-$j$ ($j=$~jet) channel for search purposes, with mono-$\gamma$, -$W^\pm$ and -$Z$ aiding most for diagnostic tasks. Herein, the keyword `mono' refers to the fact that nothing but the probe appears in the detector, so that missing transverse energy is measured alongside it.
In refs.~\cite{Abdallah:2015hma,Abdallah:2015uba}, it was pointed out that, even when the DM candidate is the same in both models\footnote{This is typically the lightest neutralino, ${\tilde{\chi}_1^0}$, which is also the Lightest Supersymmetry Particle (LSP).}, 
 the typical topologies of these processes can be very different between the MSSM and the BLSSM. This is due the fact that the mediator of DM pair production in the MSSM is an off-shell $Z$ boson while in the BLSSM can naturally be a rather massive $Z'$ boson (in the few TeV range). 
The peculiarity of the $Z'$ signal decaying invisibly (directly into DM or else via heavy (s)neutrinos in turn yielding the LSPs and light neutrinos), with respect to the $Z$ one (decaying directly into two lightest neutralinos),
 is that the final state mono-probe carries a very large (transverse) missing energy. Under these circumstances the efficiency in accessing the invisible final state and rejecting the Standard Model (SM) background is very high
altogether compensating initially smaller production rates with respect to the $Z$ case. Exploiting this feature, it has been shown that significant sensitivity exists already after 300~fb$^{-1}$ during Run 2, to
the extent that mono-$j$ events can be readily accessible at the LHC, so as to enable one to claim a prompt discovery, while mono-$\gamma$ as well as -$Z$ signals can be used simultaneously as diagnostic tools of the underlying scenario.

The recent discovery of a SM Higgs boson $h$ has however paved the way to another signal in the above category, the so-called mono-$h$ one (i.e., a mono-Higgs type) \cite{monoHreview,mono-review}. The latter is not just another probe similar to the existing ones though.
There is in fact a key difference between mono-$h$ and other mono-type searches.  In proton-proton collisions, a $j$/$\gamma$/$W^\pm$/$Z$ can be emitted directly from a light quark as Initial State Radiation (ISR) through the usual SM gauge interactions, or it may be emitted as part of the remainder of the process. 
In contrast,  ISR induced by Higgs-strahlung is highly suppressed due to the small coupling of the Higgs boson to light quarks.  Hence, unlike other mono-type signatures, 
a  mono-$h$ signal would probe exclusively the properties of the mediator and/or DM.

It is the purpose of this paper to study the scope afforded by potential mono-$h$ signals at the LHC in the BLSSM by exploiting the fact that the $h$ state can be emitted by massive objects, like the $Z'$ or even an heavy Higgs boson $h'$, both of which can couple strongly to initial state quarks and gluons, respectively. Ideally, the mono-$h$ signal to be considered here 
within the BLSSM would benefit
from the same kinematic features discussed above for the case of the other mono-types, thereby offering the twofold opportunity of at the same time establishing a signal of SUSY DM and characterising it as being incompatible with the MSSM. In particular, we will 
consider the following mono-$h$ signals: $q\bar q\to Z'\to Z(\to\nu\bar\nu)h$  and $gg\to h'\to h(\to{\tilde{\chi}_1^0}{\tilde{\chi}_1^{0*}} )h$, wherein the mono-$h$ probe eventually decays via $h\to\gamma\gamma$ and $h\to ZZ^* \to 4 l$.

The plan of our paper is as follows. The next section will be devoted to describe mono-$h$ signals arising in $B-L$ SUSY models while the one after will present the results of
our numerical analysis.  In section \ref{sec:con}, we conclude.

\section{Mono-Higgs in $B-L$ SUSY models}

In discussing mono-$h$ signals in the BLSSM, {it is useful to recall the structure of its $Z'$, Higgs  and DM sectors. }

\subsection{The $Z'$  sector in the BLSSM}

The $U(1)_Y$ and $U(1)_{B-L}$ gauge kinetic mixing can be absorbed in the covariant derivative redefinition, where the gauge coupling matrix will be transformed as follows:
\be
G = \left(\begin{array}{cc}
  g_{_{YY}} & g_{_{YB}}\\
  g_{_{BY}} & g_{_{BB}}\\
\end{array}%
\right) ~~ \Longrightarrow ~~ \tilde{G} = \left(\begin{array}{cc}
  g_1& \tilde{g}\\
  0 & g_{_{B-L}}\\
\end{array}%
\right) , %
\ee
where 
\be 
g_1 = \frac{g_{_{YY}} g_{_{BB}} - g_{_{YB}} g_{_{BY}}}{\sqrt{g_{_{BB}}^2 + g_{_{BY}}^2}},~~
g_{_{B-L}} =\sqrt{g_{_{BB}}^2 + g_{_{BY}}^2},~~
\tilde{g} = \frac{g_{_{YB}} g_{_{BB}} + g_{_{BY}} g_{_{YY}}}{\sqrt{g_{_{BB}}^2 + g_{_{BY}}^2}}.
\ee
 In this basis, one finds
\be
M_Z^2 = \frac{1}{4} (g_1^2 +g_2^2) v^2,  ~~~~ M_{Z'}^2 = g_{_{B-L}}^2 v'^2 + \frac{1}{4} \tilde{g}^2 v^2 .
\ee
Furthermore, the mixing angle between $Z$ and $Z'$ is given by 
\be 
\tan 2 \theta' = \frac{2 \tilde{g}\sqrt{g_1^2+g_2^2}}{\tilde{g}^2 + 4 (\frac{v'}{v})^2 g_{_{B-L}}^2 -g_2^2 -g_1^2}.
\ee

\subsection{The Higgs sector in the BLSSM}

The superpotential of the BLSSM is given by
\bea
\hat{W} = Y_u \hat{Q}\hat{H}_2\hat{U}^c + Y_d\hat{Q}\hat{H}_1\hat{D}^c + Y_e\hat{L}\hat{H}_1\hat{E}^c +Y_{\nu}\hat{L}\hat{H}_2\hat{N}^c
+ Y_N\hat{N}^c\hat{\eta}_1\hat{N}^c+\mu \hat{H}_1\hat{H}_2 + \mu'\hat{\eta}_1\hat{\eta}_2,\nonumber
\eea
and the soft SUSY breaking terms are given by
\begin{eqnarray}
- {\cal L}_{soft} &=& m^2_{\tilde{q}ij} \tilde{q}^*_i \tilde{q}_j + m^2_{\tilde{u}ij} \tilde{u}^*_i \tilde{u}_j + m^2_{\tilde{d}ij} \tilde{d}^*_i \tilde{d}_j
+ m^2_{\tilde{l}ij} \tilde{l}^*_i \tilde{l}_j+ m^2_{\tilde{e}ij} \tilde{e}^*_i \tilde{e}_j + m^2_{H_2} \vert H_2 \vert^2 
+ m^2_{H_1} \vert H_1 \vert^2 
\nonumber\\ 
&+&m_{\tilde{N}ij}^{2}{\tilde{N}}_{i}^{c*
}{\tilde{N}}_{j}^{c} + m^2_{\eta_1} \vert{\eta_1}\vert^2 +
 m^2_{\eta_2}\vert{\eta_2}\vert^2 +\bigg{[} Y_{u ij}^{A} \tilde{q}_i \tilde{u}_j H_2 +Y_{d ij}^{A} \tilde{q}_i \tilde{d}_j H_1 + Y_{e ij}^{A} \tilde{l}_i \tilde{e}_j H_1
\nonumber\\
 &+& \left. Y_{\nu ij}^{A}{\tilde{L}}_{i}
{\tilde{N}^c}_{j}H_2 + Y_{N ij}^{A}{\tilde{N}}_i^{c}
{\tilde{N}}_j^{c}\eta_{1} +B \mu H_2 H_1+ B \mu^\prime \eta_1 \eta_2 +
\frac{1}{2} M_a \lambda^a \lambda^a + M_{BB'} \tilde{B} \tilde{B'} + h.c.
\right],\nonumber %
\label{Lsoft}%
\end{eqnarray}%
where $(Y_f^A)_{ij} = (Y_f)_{ij} A_{ij}$, the tilde denotes the scalar components of the chiral superfields as well as the fermionic components of the vector superfields and $\lambda^a$ are fermionic components of  the vector superfields. 
The
Vacuum Expectation Values
 (VEVs) of the Higgs fields are given by $ \langle{\rm Re} H_i^0\rangle=v_i/\sqrt{2}$ and $\langle{\rm Re} \eta^0_i\rangle=v'_i/\sqrt{2}$.
To obtain the masses of the physical neutral Higgs bosons, one makes the usual redefinition of the Higgs fields, i.e.,
$H_{1,2}^0 = (v_{1,2} + \sigma_{1,2} + i \phi_{1,2})/\sqrt{2} $ and 
$\eta_{1,2}^0 =(v'_{1,2} + \sigma'_{1,2}  + i \phi'_{1,2})/\sqrt{2}$,
where $\sigma_{1,2}= {\rm Re} H_{1,2}^0$, $\phi_{1,2}={\rm Im} H_{1,2}^0$, $\sigma'_{1,2}= {\rm Re} \eta_{1,2}^0$ and $\phi'_{1,2}={\rm Im} \eta_{1,2}^0 $. The real parts correspond to the CP-even Higgs bosons and the imaginary parts correspond to the CP-odd Higgs bosons. 
The mass of the BLSSM-like CP-odd Higgs $A'$ is given by 
\be
m_{A'}^{{2}} =\frac{2 B {\mu'}}{\sin2 \beta'} \sim {\cal O}(1~{\rm TeV}),
\ee 
whereas those of the BLSSM CP-even neutral Higgs fields, at tree level, are given by
\bea
{m}^2_{h',H'} = \frac{1}{2} \Big[ ( m^2_{A'} + M_{{Z'}}^2 ) \mp \sqrt{ ( m^2_{A'} + M_{{Z'}}^2 )^2 - 4 m^2_{A'} M_{{Z'}}^2 \cos^2 2\beta' }\;\Big].
\eea
If $\cos^2{{2}\beta'} \ll 1$, one finds that the lightest $B-L$ neutral Higgs mass is given by %
\be%
{m}_{h'}\; {\simeq}\; \left(\frac{m^2_{A'} M_{{Z'}}^2 \cos^2 2\beta'}{{m^2_{A'}+M_{{Z'}}^2}}\right)^{\frac{1}{2}} \simeq {\cal O}(100~ {\rm GeV}).%
\ee%

\subsection{DM in the BLSSM}

Now, we consider the neutralino sector in the BLSSM. The  neutral
gaugino-higgsino mass matrix can be written as \cite{Khalil:2007dr}: %
\bea
&&{\cal M}_7({\tilde B},~{\tilde W}^3,~{\tilde
H}^0_1,~{\tilde H}^0_2,~{\tilde B'},~{\tilde \eta_1},~{\tilde
\eta_2}) \equiv \left(\begin{array}{cc}
{\cal M}_4 & {\cal O}\\
 {\cal O}^T &  {\cal M}_3\\
\end{array}\right),
\eea%
where the ${\cal M}_4$ is the MSSM-type neutralino mass matrix \cite{Haber:1984rc,Gunion:1984yn,ElKheishen:1992yv,Guchait:1991ia} and
${\cal M}_{3}$ is $3\times 3$ additional neutralino mass matrix,
which is given by%
\be%
{\cal M}_3 = \left(\begin{array}{ccc}
M_{B'} & -g_{_{B-L}}v'_1  & g_{_{B-L}}v'_2 \\
-g_{_{B-L}}v'_1 & 0 & -\mu'  \\
g_{_{B-L}}v'_2 & -\mu' & 0\\
\end{array}\right).
\label{mass-matrix.1} \ee
In addition, the off-diagonal matrix ${\cal O}$ is given by
\be%
{\cal O} = \left(\begin{array}{ccc}
\frac{1}{2}M_{BB'} &~~~0~~~& 0 \\
0 & 0 & 0  \\
-\frac{1}{2}\tilde{g}v_1 &~~~0~~~& 0\\
\frac{1}{2}\tilde{g}v_2&~~~0~~~&0\\
\end{array}\right).
\label{mass-matrix.1} \ee
Note that these off-diagonal elements vanish identically if $\tilde{g}=0$. In this case, one diagonalises the real matrix ${\cal M}_{7}$ with
a symmetric mixing matrix $V$ such as
\be V{\cal
M}_7V^{T}={\rm diag}(m_{\tilde\chi^0_k}),~~k=1,\dots,7.\label{general} \ee In
these conditions, the LSP has the following decomposition 
\be 
\tilde\chi^0_1=V_{11}{\tilde B}+V_{12}{\tilde
W}^3+V_{13}{\tilde H}^0_1+V_{14}{\tilde
H}^0_2+V_{15}{\tilde B'}+V_{16}{\tilde \eta_1}+V_{17}{\tilde
\eta_2}. 
\ee 
The LSP is called pure
$\tilde B'$ if $V_{15}\sim1$ and $V_{1i}\sim0$ for $i\neq5$, and pure $\tilde\eta_{1(2)}$ if $V_{16(7)}\sim1$
and all the other coefficients are close to zero. 

\subsection{Mono-Higgs channels in SUSY models}
\label{sec:MonoH-MSSM}
In discussing mono-$h$ signals in the BLSSM, it is useful to contrast their dynamics against that of the MSSM, for
which several analyses already exist \cite{monoHreview}. In the MSSM, where the DM particle is the lightest neutralino, $\tilde{\chi}_1^0$, just like
in our BLSSM construction, 
we have three classes of mono-Higgs channels, to which we will dedicate three separate subsections below\footnote{Note that in the BLSSM versus MSSM comparison we neglect topologies
where a $h'/H'/A'$ is produced in place of the SM-like state and cascade down to it (invisibly for the rest of the event).}.

\subsubsection{Mono-Higgs as a final state}
\begin{figure}[!t]
\begin{center}
\begin{tabular}{cc}
\multirow{-5}{*}{\includegraphics[width=0.3\textwidth]{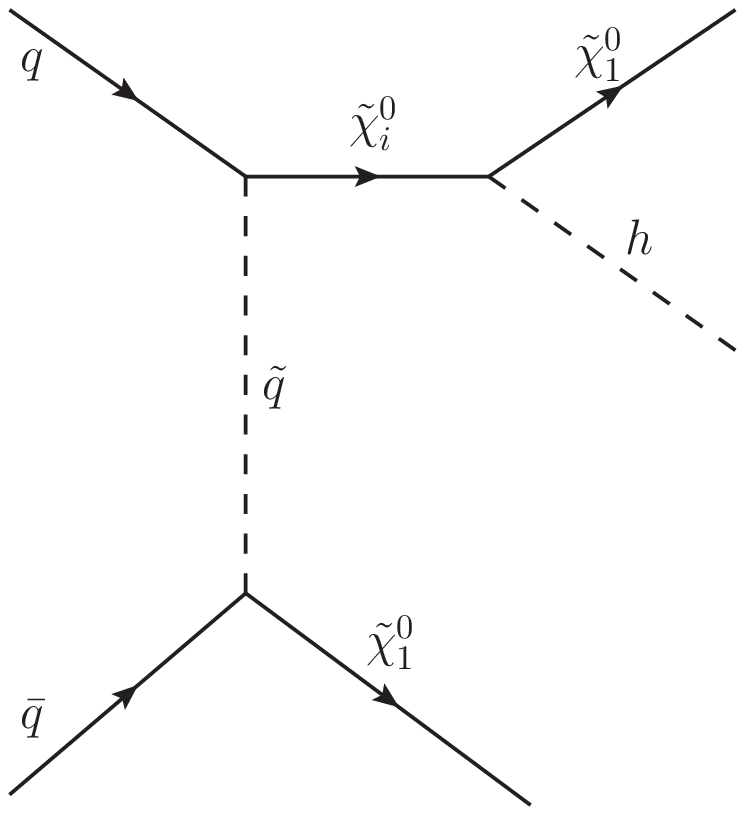}} &~~~\includegraphics[width=0.35\textwidth]{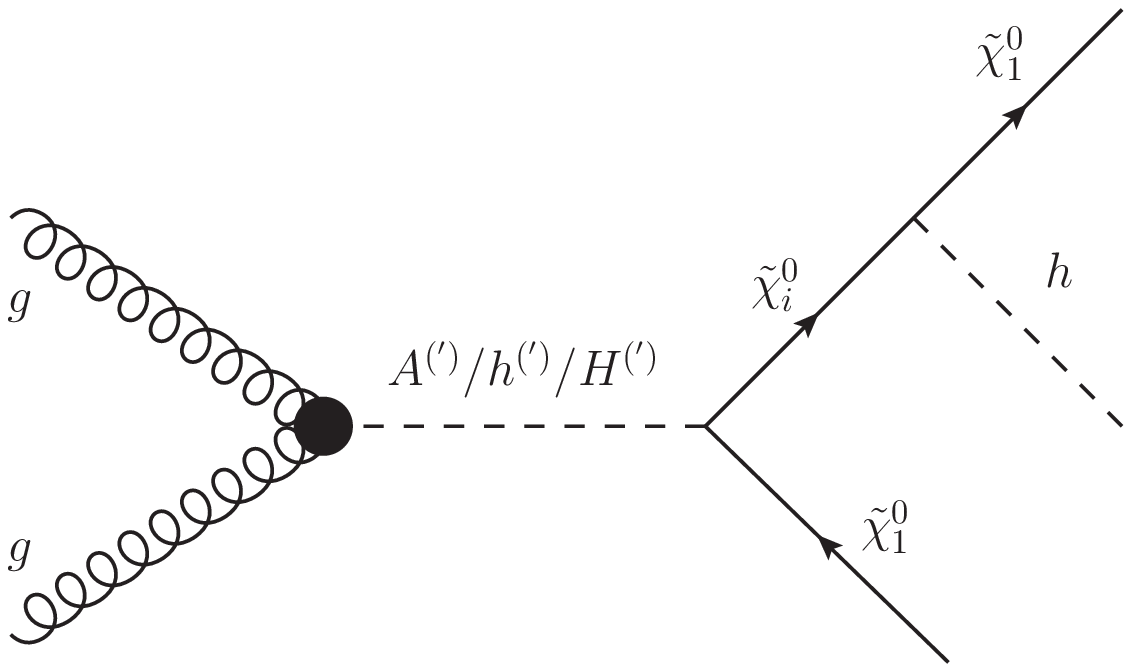}\\
&~~~\includegraphics[width=0.3\textwidth]{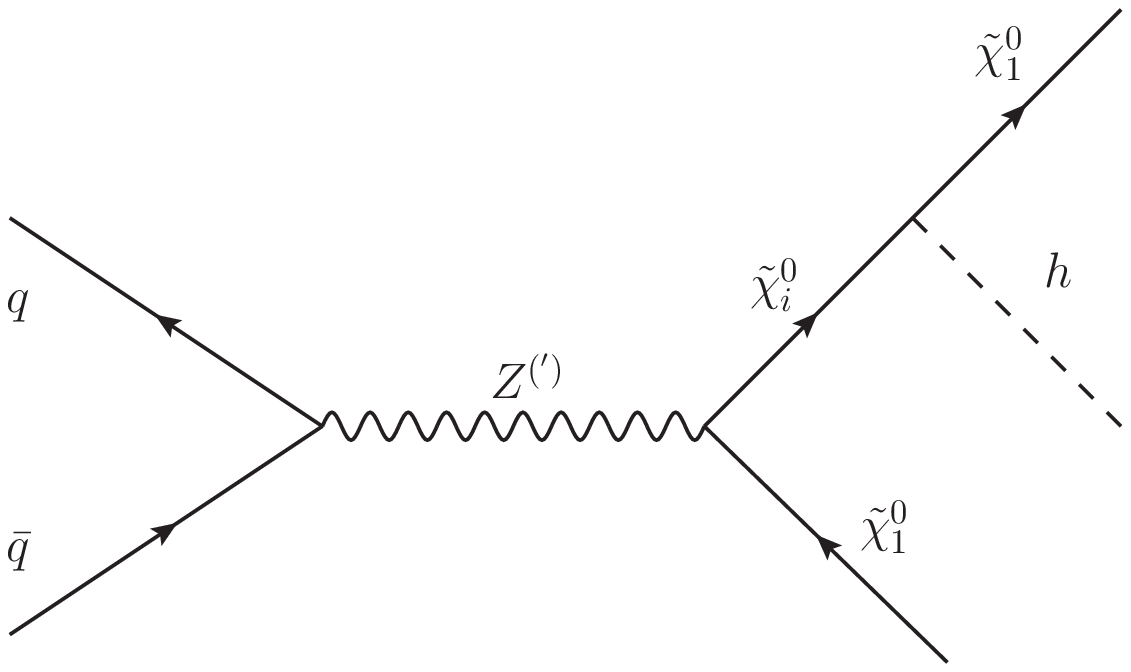}
\end{tabular}
\caption{Mono-Higgs as a final state: $q\bar{q}\to\tilde{\chi}_1^0\tilde{\chi}_i^0\to\tilde{\chi}_1^0\tilde{\chi}_1^0 h$ with $\tilde{q}$ exchange
(left diagram),
$gg \to A^{(')}/h^{(')}/H^{(')}\to\tilde{\chi}_1^0\tilde{\chi}_i^0\to\tilde{\chi}_1^0\tilde{\chi}_1^0 h$ (top-right diagram) and $q\bar{q} \to Z^{(')}\to\tilde{\chi}_1^0\tilde{\chi}_i^0\to\tilde{\chi}_1^0\tilde{\chi}_1^0 h$ (bottom-right diagram).}
\label{fig:FSR_MSSM}
\end{center}
\end{figure}

In this class, we have three types of MSSM channels, that we can group in two subsets depending on the mediators, see figure \ref{fig:FSR_MSSM}: ($i$) $q\bar{q}\to\tilde{\chi}_1^0\tilde{\chi}_i^0\to\tilde{\chi}_1^0\tilde{\chi}_1^0 h$ with $\tilde{q}$ exchange, the typical value of the cross section of this channel being of order ${\cal{O}}(10^{-7})$~pb and it is worth to note that it comes from a large $\tilde{q}$ mass; ($ii$) $gg \to A/h/H\to\tilde{\chi}_1^0\tilde{\chi}_i^0\to\tilde{\chi}_1^0\tilde{\chi}_1^0 h$ and $q\bar{q} \to Z\to\tilde{\chi}_1^0\tilde{\chi}_i^0\to\tilde{\chi}_1^0\tilde{\chi}_1^0 h$, where $i=2,3,4$, again, the typical value of the cross section of these channels being of order ${\cal{O}}(10^{-7})$~pb  (in case of $A \ \text{and} \ H$ mediators, these channels are suppressed due to their small  production rates as well as the off-shell decay $\tilde{\chi}_i^0\to\tilde{\chi}_1^0 h$, while in the case of $h \ \text{and} \ Z$ mediators, although they have larger production rates, the suppression coming from their off-shell decays is substantial).      

The BLSSM can add a few contributions to these topologies (specifically, to the two graphs on the right-hand side of
 figure \ref{fig:FSR_MSSM}). Wherever a $Z$ is present in the MSSM, a $Z'$ can also
contribute. Furthermore, for each of the neutral MSSM Higgs states, $h,~H$ and $A$, there corresponds in the BLSSM a
primed version, $h',~H'$ and $A'$, wherein the $h'$ can have a mass similar to the $h$ one (i.e., just above 125 GeV)
while the other two states are generally much heavier \cite{Abdallah:2014fra,Hammad:2015eca,Khalil:2015vpa,Hammad:2016trm}, most
notably in its inverse seesaw version \cite{Khalil:2015naa}. Hence, the potential to increase the sensitivity of experimental
analyses is twofold. On the one hand, the $Z'$ can be produced resonantly as its current mass limits within the BLSSM enable on-shell
decays $Z'\to \tilde{\chi}_1^0\tilde{\chi}_i^0\to\tilde{\chi}_1^0\tilde{\chi}_1^0 h$, where $i=2,\dots,7$. On the other hand, $h'$ can also be resonant in regions of parameter space where $m_{h'}>m_h+2 m_{\tilde{\chi}_1^0}$, which are indeed presently accessible within the BLSSM.

\subsubsection{Mono-Higgs as an intermediate state}
\begin{figure}[!t]
\begin{center}
\begin{tabular}{cc}
\multirow{-5}{*}{\includegraphics[width=0.25\textwidth]{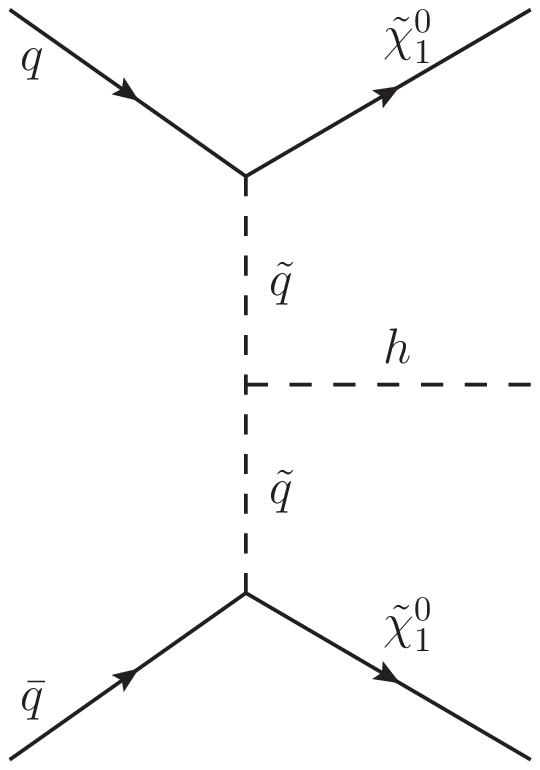}} & ~\includegraphics[width=0.35\textwidth]{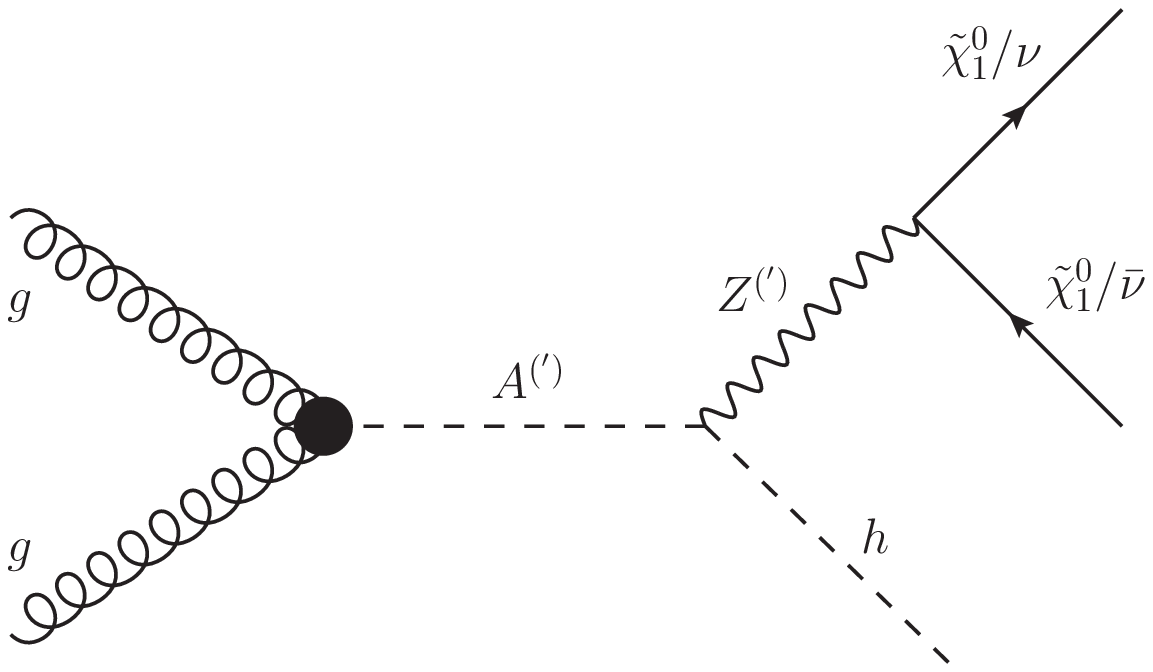}~\includegraphics[width=0.35\textwidth]{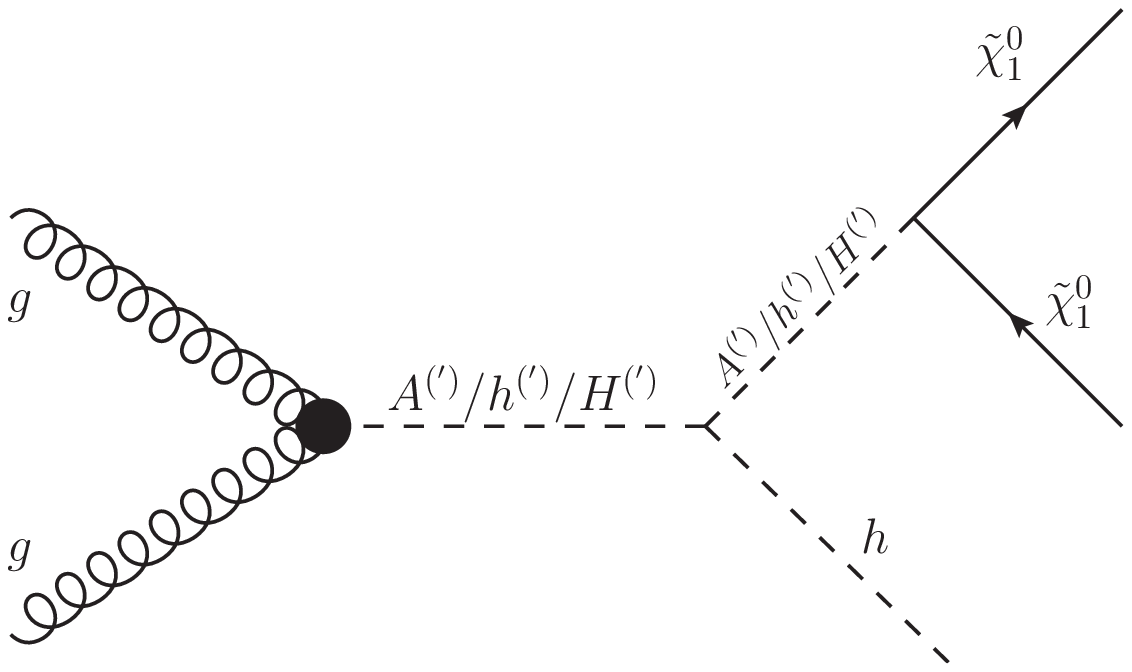}\\
&~\includegraphics[width=0.35\textwidth]{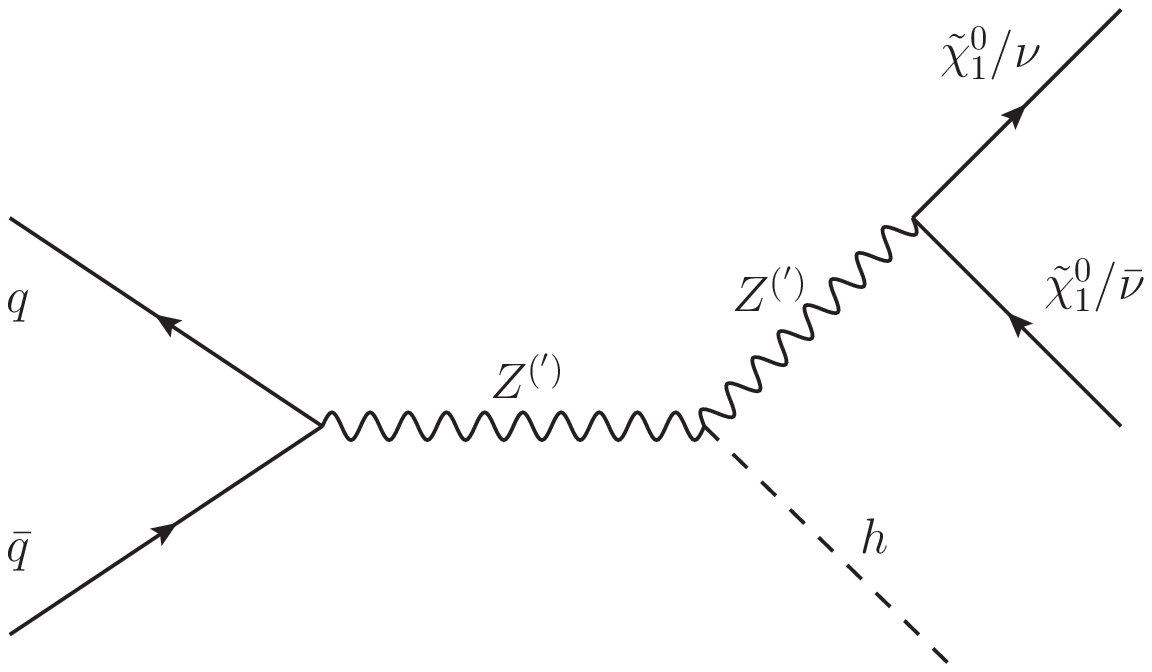}~\includegraphics[width=0.35\textwidth]{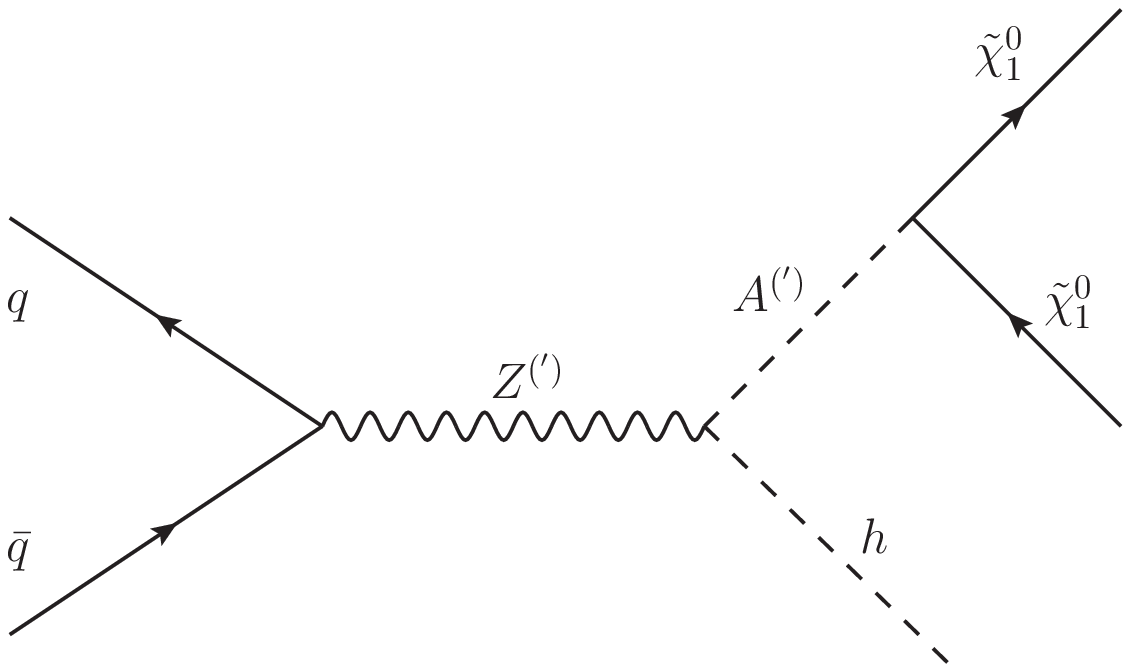}
\end{tabular}
\caption{Mono-Higgs as an intermediate state: $qq\to\tilde{\chi}_1^0\tilde{\chi}_1^0 h$ with $\tilde{q}\tilde{q}$ exchange (left diagram), $gg\to A^{(')}\to Z^{(')} h\to\tilde{\chi}_1^0\tilde{\chi}_1^0 h/\nu\bar{\nu}h$ plus $q\bar{q}\to Z^{(')} \to Z^{(')} h\to\tilde{\chi}_1^0\tilde{\chi}_1^0 h/\nu\bar{\nu}h$ (center diagrams)
and
 $gg\to A^{(')}/h^{(')}/H^{(')} \to A^{(')}/h^{(')}/H^{(')}~h\to\tilde{\chi}_1^0\tilde{\chi}_1^0 h$ and $q\bar{q}\to Z^{(')} \to A^{(')} h\to\tilde{\chi}_1^0\tilde{\chi}_1^0 h$ (right diagrams).}
\label{fig:INSR_MSSM}
\end{center}
\end{figure}

In this class, we have five types of MSSM channels, that we can group in three subsets depending on the mediators, see figure \ref{fig:INSR_MSSM}: ($i$) $qq\to\tilde{\chi}_1^0\tilde{\chi}_1^0 h$ with $\tilde{q}\tilde{q}$ exchange, its typical cross section 
being of order ${\cal{O}}(10^{-6})$~pb, again, this suppression stands from the large $\tilde{q}$ mass; ($ii$) $gg\to A\to Z h\to\tilde{\chi}_1^0\tilde{\chi}_1^0 h$ (its cross section being of ${\cal{O}}(10^{-5})$~pb  due to the smallness of the production 
rates of the $A$) and $q\bar{q}\to Z \to Z^* h\to\tilde{\chi}_1^0\tilde{\chi}_1^0 h$ (its cross section being very suppressed due to the off-shell decay of the $Z$); ($iii$) $gg\to A/h/H \to A/h/H~h\to\tilde{\chi}_1^0\tilde{\chi}_1^0 h$ (its cross section being of ${\cal{O}}(10^{-4})$~pb
owing to the dominant channel $H\to h h$) and $q\bar{q}\to Z \to A h\to\tilde{\chi}_1^0\tilde{\chi}_1^0 h$ (its cross section being very suppressed due to the off-shell decay of the $Z$).

Within the BLSSM, again, wherever a $Z$ is involved a $Z'$ also is and, likewise, wherever a $h/H/A$ enters also a
$h^{(')}/H^{(')}/A^{(')}$ appears (this is limited to the center and right topologies in figure \ref{fig:INSR_MSSM}).
Like previously, we expect some tangible contribution of specific BLSSM nature whenever the (heavy) $Z'$ and/or (light) $h'$ can resonate, so long that no heavy $H'$ and $A'$ states are present in the same channel.

\subsubsection{Mono-Higgs as an initial state}
\begin{figure}[!t]
\begin{center}
\includegraphics[width=0.25\textwidth]{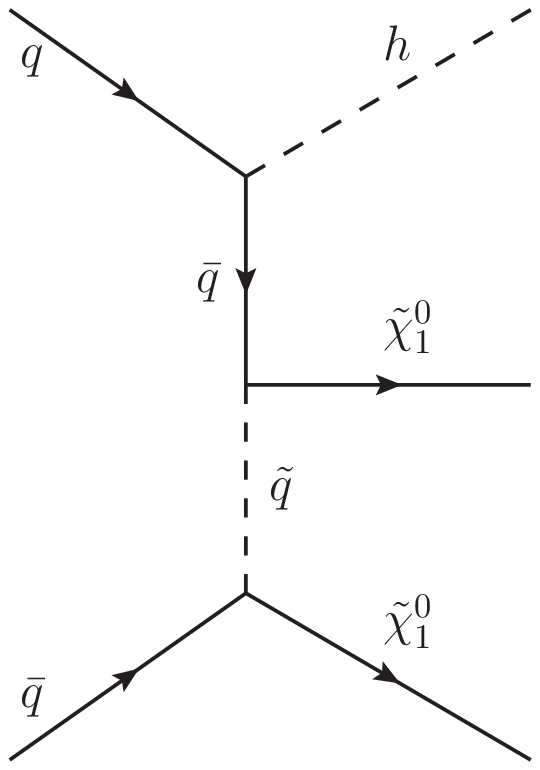}~\includegraphics[width=0.33\textwidth]{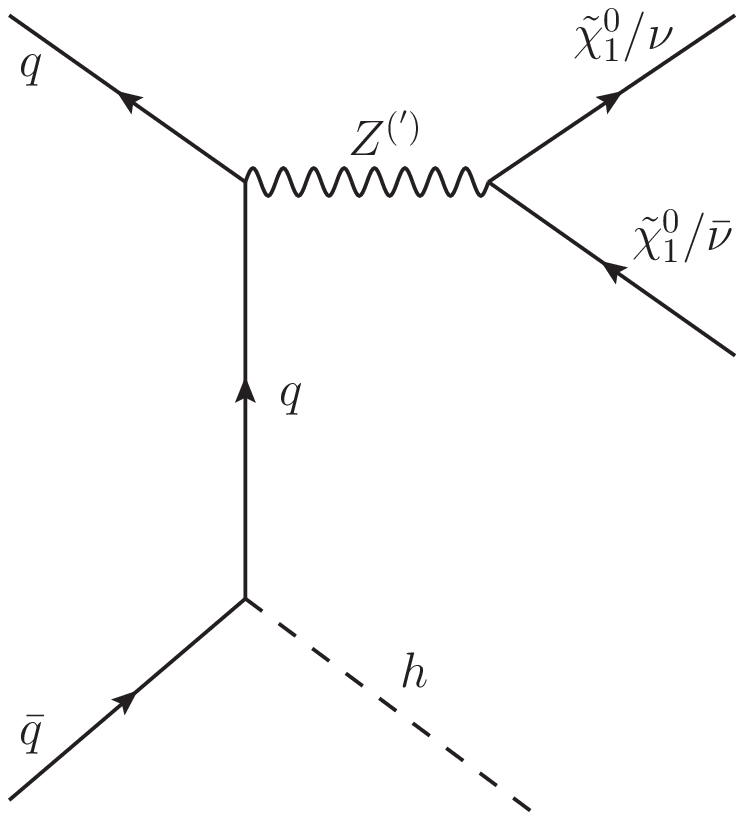}~\includegraphics[width=0.33\textwidth]{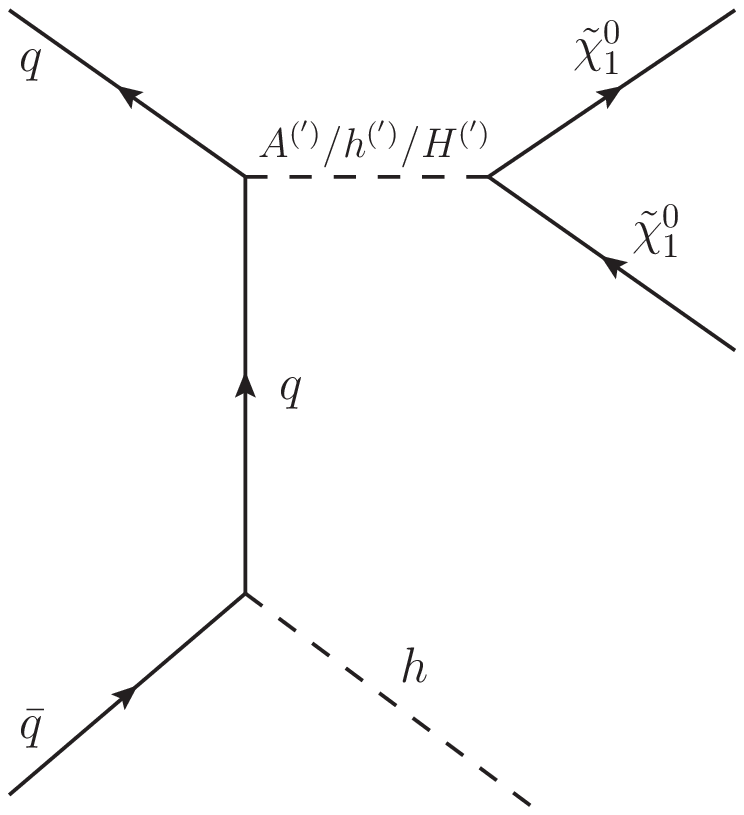}   
\caption{Mono-Higgs as an initial state: $q\bar{q}\to\tilde{\chi}_1^0\tilde{\chi}_1^0 h$ with $q\tilde{q}$ exchange (left diagram), $q\bar{q}\to Z^{(')}h\to\tilde{\chi}_1^0\tilde{\chi}_1^0 h/\nu\bar{\nu}h$ with $q$ exchange (center diagram) and $q\bar{q}\to A^{(')}/h^{(')}/H^{(')}~h\to\tilde{\chi}_1^0\tilde{\chi}_1^0 h$ with $q$ exchange, respectively.}
\label{fig:ISR_MSSM}
\end{center}
\end{figure}

In this class, we have three types of MSSM channels, each characterised by its on specific mediators, 
see figure \ref{fig:ISR_MSSM}: ($i$) $q\bar{q}\to\tilde{\chi}_1^0\tilde{\chi}_1^0 h$ with $q\tilde{q}$ exchange; ($ii$) $q\bar{q}\to Zh\to\tilde{\chi}_1^0\tilde{\chi}_1^0 h$ with $q$ exchange; ($iii$) $q\bar{q}\to A/h/H~h\to\tilde{\chi}_1^0\tilde{\chi}_1^0 h$ with $q$ exchange. The cross sections of all these types are very suppressed due to a very small coupling of $h$ with $q\bar{q}$. Moreover, in the first mode, one has the additional depletion induced by a large $\tilde{q}$ mass. 

In this case the BLSSM has little to add to the MSSM. The only possible enhancement to the overall rate could come from
$Z'$ exchange in the center diagram of figure \ref{fig:ISR_MSSM} when the graph resonates, as in the left topology
there is no difference while in the right one additional $h'/H'/A'$ states are more suppressed than their un-primed versions.

\subsection{Analysis strategy for mono-Higgs searches in the BLSSM}
\label{sec:analysis}

We initially detail the BLSSM parameter space tested, by delineating the intervals used to sample
the independent parameters of this scenario assuming a low energy scale construction, then we explain in  detail the numerical procedure used for this analysis, where simulated events for Signal ($S$) and Background ($B$) were generated 
through a standard sequence of a matrix element calculator, a  Monte Carlo (MC)  program and  LHC
detector software. In the two following subsections we explain how to extract di-photon and four-lepton signatures for
our mono-$h$ probe, mediated by either $Z'$ or $h'$ intermediate production as, following the discussions in the previous section, these are the distinctive features of the BLSSM versus the MSSM.  

\subsubsection{LHC current exclusion and parameter space}

A summary of the parameter space points used for all the signals considered here is reported in  table \ref{table:param}. The inputs  used for simulating a SM-like Higgs boson
produced 
 in association with low missing transverse energy through an $h'$ mediator are presented in the first three rows while those for
emulating a  SM-like Higgs boson produced in association with high missing transverse energy mediated by a $Z'$ are presented in the last row.
{\small
\begin{table}[h!]
\begin{center}
\begin{tabular}{|c|c|c|c|c|c|c|c|c|}
\hline
$M_{Z'}\,[\text{GeV}]$&$g_{_{B-L}}$&$\tilde{g}$&$\theta'$&$m_{h'}\,[\text{GeV}]$&$m_{\tilde{\chi}^\pm_1}\,[\text{GeV}]$&$m_{\tilde{g}}\,[\text{GeV}]$&$m_{\tilde{\chi}^0_{1}}\,[\text{GeV}]$\\
\hline     
1916.5&0.27& $-0.79$& $1.8\times 10^{-3}$&265.3&772&6178&10.6\\
\hline     
1645.1&0.23& $-0.89$& $2.7\times 10^{-3}$&264.7&771&6178&29.9\\
\hline     
1468.4&0.21& $-0.89$& $3.4\times 10^{-3}$&279.1&619&6185&48.6\\
\hline\hline
2396.5&0.40& $-0.47$& $8.2\times 10^{-4}$&332.6&920&6198&412\\ 
\hline
\end{tabular}
\caption{The first three benchmark points (rows) for the $h'$ mediated signal and the last one for the $Z'$ mediated signal.}\label{table:param}
\end{center}
\end{table}
}

As intimated, we will study here  the decay of heavy boson $Z'$ and light scalar $h'$ mediators to SM-like Higgs and some  amount of missing traverse energy, where the SM-like Higgs boson decays to a $4l$ (electrons and muons only) or $\gamma\gamma$ final state. The presence of missing transverse energy $\met$ in  the event is one of main distinguishing characteristic of the signal, which is defined as the negative sum of the transverse momenta of all reconstructed objects. Thus, it depends on the reconstruction of all charged particles, especially jets which can be responsible for inducing unwanted amounts of $\met$. Another variable useful to reduce the background further and enhance the signal is the transverse mass $M_T$ of the four-lepton and di-photon systems defined as follows:
\begin{equation}\label{eq:MT}
M_T^2(f) = \left(\sqrt{M^2(f)+p^2_T(f)}+\left|p_T^{\text{miss}}\right|\right)^2-\Big[\vec{p}_T(f)+ \vec{p}_T^{\text{~miss}}\Big]^2,
\end{equation}
where $M(f)$ and $p_T(f)$ are the invariant mass and transverse momentum, respectively, of the final state particles which are $f=\gamma\gamma$ and $4l$. In the end, a set of standard cuts will be chosen to enhance the $S$-to-$B$ ratio ($S/B$), yet vetoing the above transverse mass range above $250$~GeV improves the latter significantly in case of low missing transverse energy induced by the $h'$ mediator, while this variable has less relevance for the case of a heavy mediator $Z'$.

It is worth to note that the spectra associated with the above mentioned benchmark points in table \ref{table:param} are consistent with  current LHC bounds. Also, for the $Z'$ mass, we assured the LEP constraints: $M_{Z'}/g_{_{B-L}}>6$~TeV and $\theta'\lesssim {\cal O}(10^{-3})$ \cite{Cacciapaglia:2006pk}. Moreover, the LSP satisfies the LUX bounds on the direct search for  DM \cite{Akerib:2015rjg} and other direct detection experimental limits \cite{Aprile:2012zx,Aprile:2012nq}. However, the relic abundance depends on the details of the underlying cosmology (thermal or non-thermal abundance) so its constraints will not be considered here \cite{Kolb,Giudice:2000ex,Moroi:1999zb,Abdallah:2015hza}.  

\subsubsection{Numerical tools}
Both signal and background are computed with MadGraph5 \cite{Madgraph5} that is used to estimate  multi-parton amplitudes and to generate events for the calculation of the cross sections as well as for subsequent processing. The production cross sections for $h'$ are calculated at Next-to-Leading Order (NLO) using an effective coupling calculated by SPheno \cite{PorodSPheno,florianSPheno} while those for $Z'$ mediation have LO normalisation. PYTHIA \cite{Sjostrand:2006za} is used for showering, hadronisation, heavy flavour decays and for adding the soft underlying event. The simulation of the response of the ATLAS and CMS detectors was done with the DELPHES package \cite{deFavereau:2013fsa}. Reconstructed objects are simulated from the parametrised detector response and includes tracks, calorimeter deposits and high level objects such as isolated electrons, jets, taus and missing transverse momentum.


\section{BLSSM signals and LHC sensitivity}

In this section, we concentrate on mono-$h$ signals which are specific to the BLSSM, i.e., those associated to $Z'$ and $h'$ induced topologies, wherein these states act as mediators for DM creation. As intimated, we shall assume the SM-like Higgs state $h$ to decay into the two channels that enable an effective Higgs mass reconstruction, so as to exploit the measured 125 GeV mass for background suppression. These are $h\to \gamma\gamma$ and $h\to ZZ^*\to 4l$, where $l=e$ or $\mu$. We shall do so in two separate subsections.  

\subsection{The $\gamma\gamma$ + $\met$ signature}
\
In this subsection we study the final state with  di-photons associated with missing transverse energy, $\met$, which  comes from neutrinos in the $Z'$ mediated channel and from neutralinos in the $h'$ mediated channel.
In this analysis, we are looking for an excess over the SM predictions in the di-photon mass spectrum after a selection in terms of the missing transverse energy and/or transverse mass. For these events, pairs of photons are reconstructed to form the SM-like Higgs boson. To enhance $S/B$ we first consider, for both  $Z'$ and $h'$ signals, the kinematic selection used in the ATLAS analysis of ref.~\cite{Aad:2015yga}, as follows.
\begin{enumerate}
\item The absolute value of the pseudo-rapidity of both photon candidates is required to be below 2.5.
\item The invariant mass $m_{\gamma\gamma}$ of the photon pair is required to be above $95$~GeV.
\item The transverse momentum $p_T$ of the leading (subleading) photon has to be above $30(20)$~GeV.
\item The $p_T / m_{\gamma\gamma}$ ratio of the leading (subleading) photon has to be above $1/3(1/4)$. 
\end{enumerate}

\begin{figure}[t!]
\begin{center}
\includegraphics[width=0.54\textwidth]{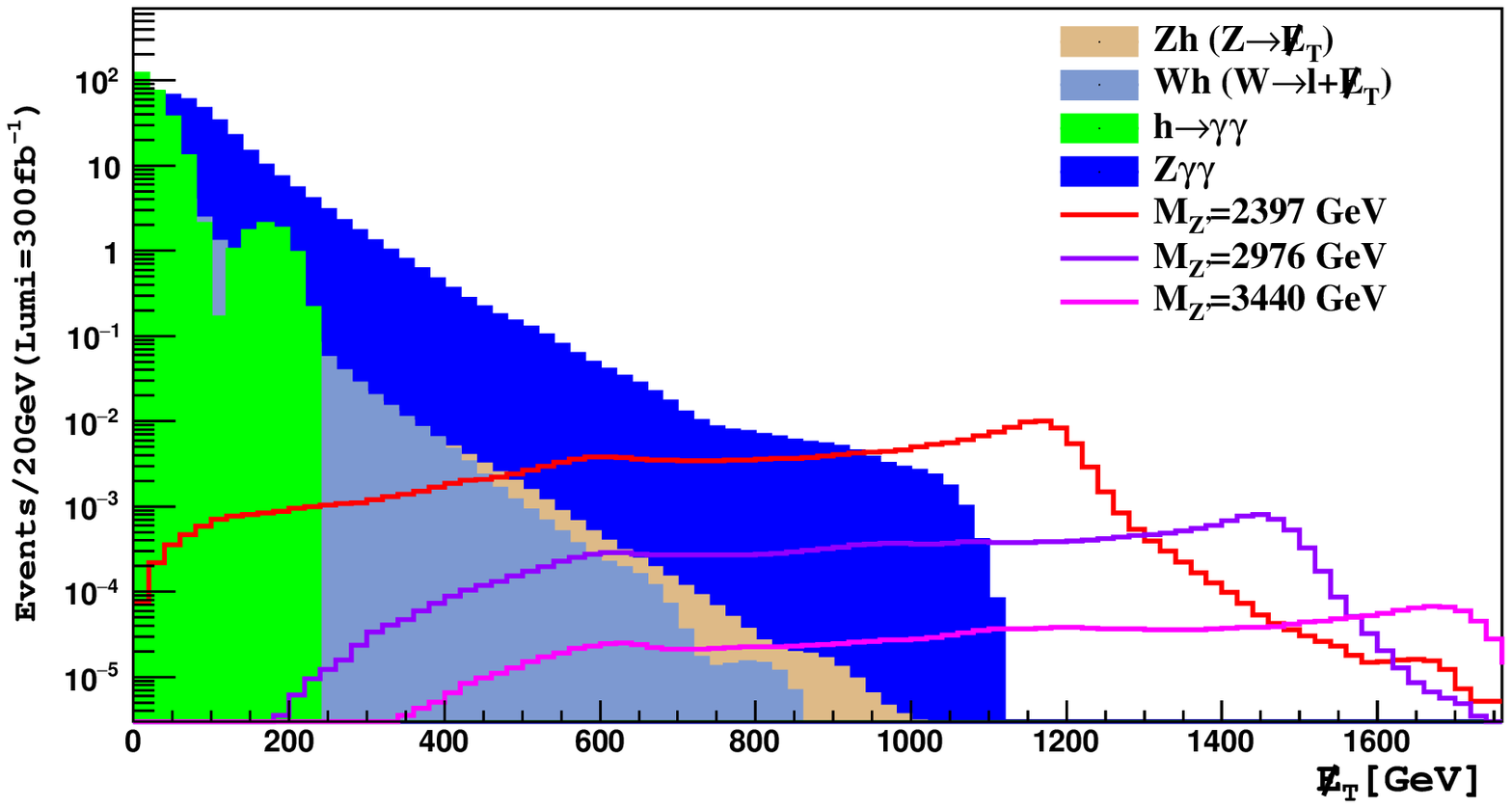}~~\includegraphics[width=0.41\textwidth]{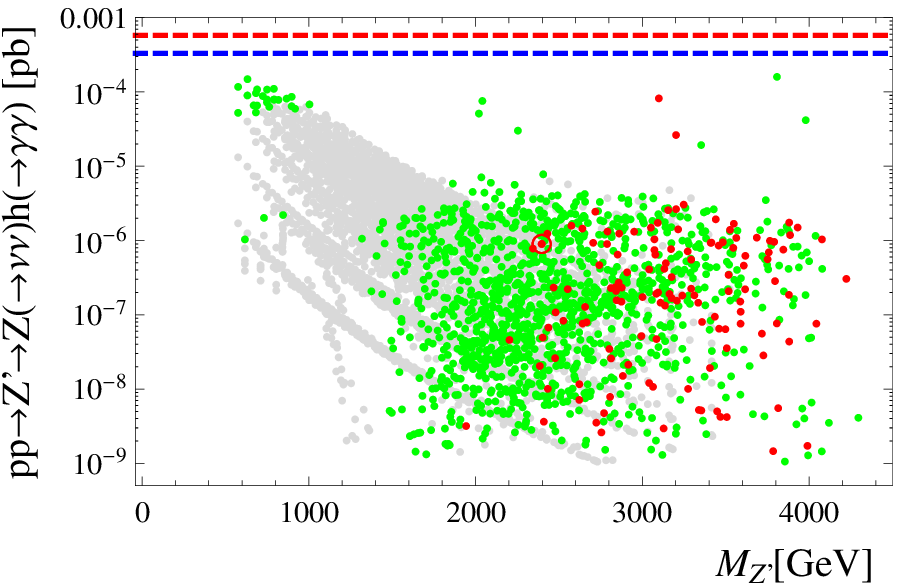}
\caption{(Left panel) Number of  events of both signal ($pp\to Z'\to Zh\to\gamma\gamma+\met$) and its
relevant backgrounds generated at $14$~TeV and normalised per bin width after $300$~fb$^{-1}$ of integrated luminosity versus $\met$ after considering all cuts applied by ATLAS \cite{Aad:2015yga}. (Right panel) The gray and green points are the signal benchmarks excluded by LEP constraints ($M_{Z'}/g_{_{B-L}}<6$~TeV and $\theta' > {\cal O}(10^{-3})$, respectively) and the red points are the allowed ones, all mapped versus the $Z'$ mass. The red circled point is the last benchmark point in table \ref{table:param}. The blue and red dashed lines are the one and two sigma exclusion limits, respectively, by ATLAS \cite{Aad:2015yga}.}
\label{fig:Zp_aa_nocut}
\end{center}
\end{figure}

\begin{figure}[h!]
\begin{center}
\includegraphics[width=0.52\textwidth]{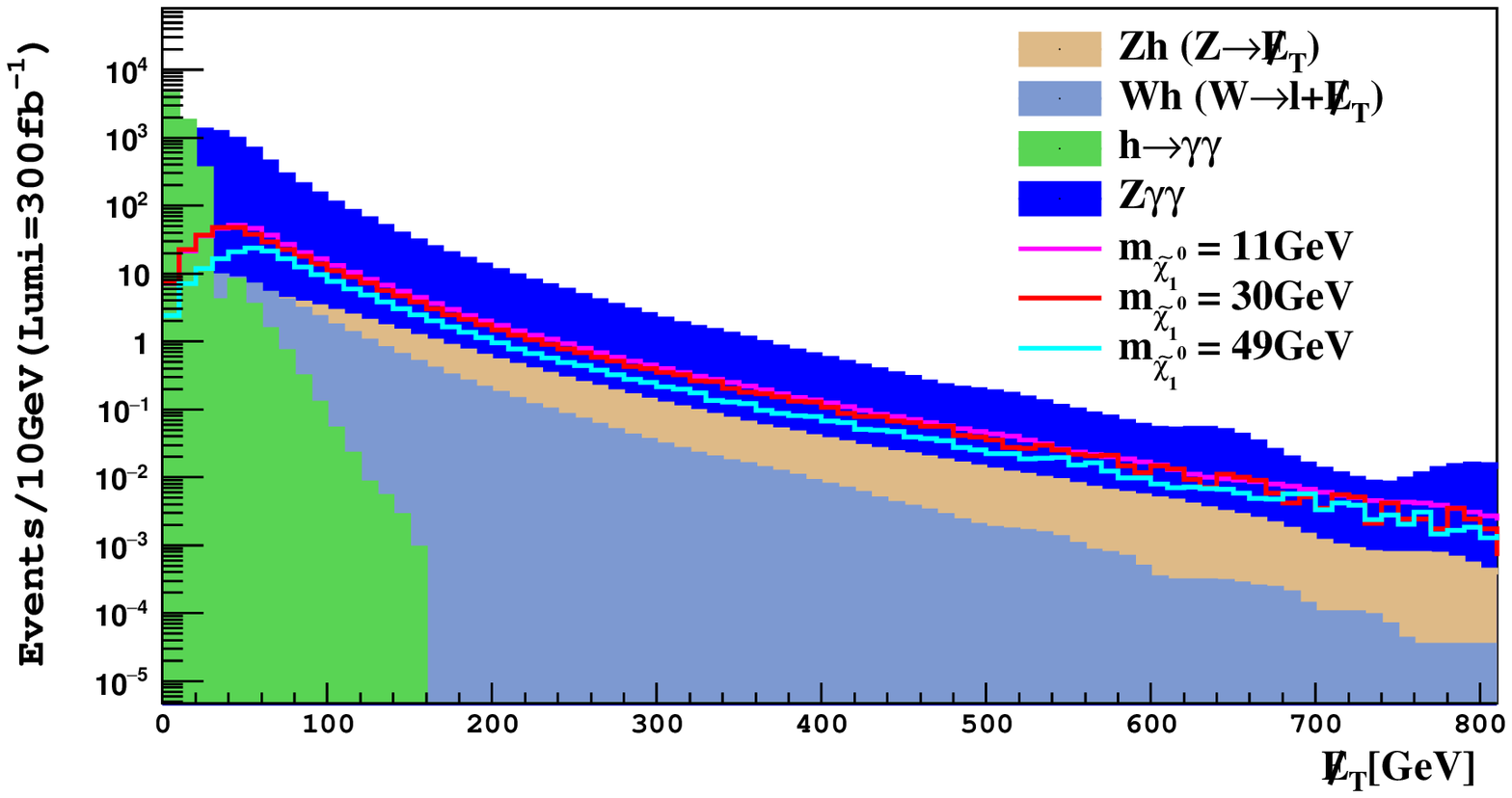}~~\includegraphics[width=0.52\textwidth]{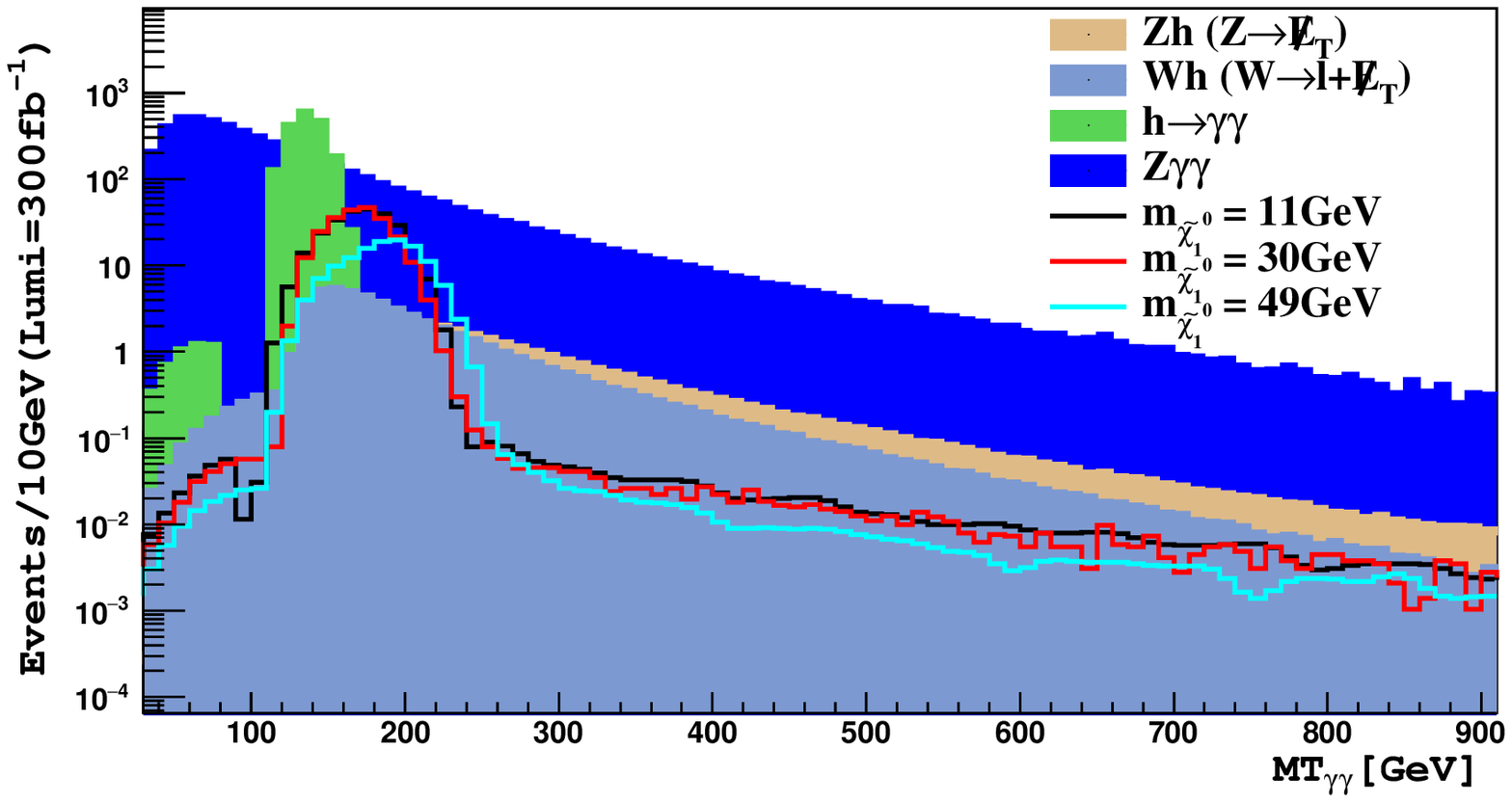}
\caption{Number of events of both signal ($pp\to h^\prime \to hh\to\gamma\gamma +\met$)  and its relevant backgrounds
generated at $14$~TeV and normalised per bin width after $300$~fb$^{-1}$ of integrated luminosity versus $\met$ (left panel)
and  $M_T({\gamma\gamma})$ (right panel)  after
  considering all cuts applied by ATLAS \cite{Aad:2015yga}.}
\label{fig:hp_aa_nocut}
\end{center}
\end{figure}

\begin{figure}[h!]
\begin{center}
\includegraphics[width=0.52\textwidth]{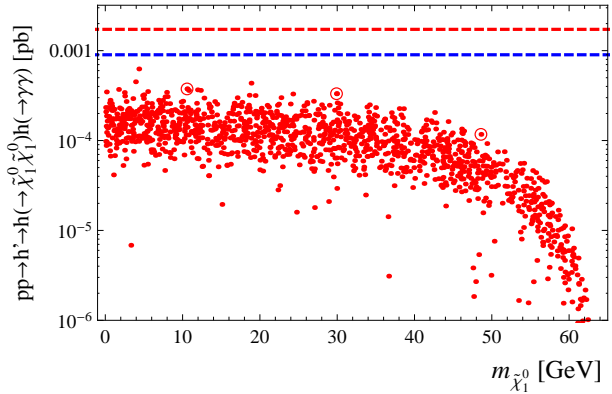}
\caption{The red points are the allowed signal benchmarks mapped versus the LSP mass. The red circled points are the first three benchmark points in table \ref{table:param}. The blue and red dashed lines are one and two sigma exclusion limits, respectively, by ATLAS \cite{Aad:2015yga}.}
\label{fig:hp1_aa_nocut}
\end{center}
\end{figure}

Owing to the difference between the two signal mediators the set of kinematic cuts used is different depending upon whether we are looking at $Z'$ or $h'$ topologies.  In case of a $Z'$ mediator the most powerful observable for suppressing the background is $\met$, which is rather large for the signal, owing to the large value of the $Z'$ mass, see the left-hand side of figure \ref{fig:Zp_aa_nocut}. In addition,  the di-photon mass spectrum characterises the signal around the the $h$ mass value, where it tends to collect owing to the underlying $h$ resonance (this also happens for the $Zh(\to\gamma\gamma)$, $W^\pm h(\to\gamma\gamma)$   and $h\to\gamma\gamma$ noises, though, but not for
the $Z\gamma\gamma$ continuum background). So, in the end, we enforce
the following selection: $\met > 550$~GeV and $110$~GeV $< m_{\gamma\gamma}<130$~GeV. (Notice that we also ought to veto against a high $p_T$ and central lepton from 
$W^\pm(\to l\nu) h(\to\gamma\gamma)$ events.)
The benefits of this approach are clearly shown in table \ref{table:zpaa} in terms of increasing substantially $S/B$. However,
it is obvious that the event rate associated to the chosen $Z'$ benchmark is too poor for this becoming a viable channel
at the LHC during its lifetime, including a high-luminosity option \cite{Gianotti:2002xx} (where the instantaneous luminosity of the LHC can be
increased up to a factor of 10). Unfortunately, the conclusion will not change if we were to choose any other benchmark from the right-hand side of figure \ref{fig:Zp_aa_nocut}.
Concerning $h'$ topologies, owing to the much lower mediator mass involved (from  $\approx260$ to $\approx280$ GeV), a (necessarily 
 low) $\met$ cut of, say, 100 GeV is not powerful to enhance $S/B$, in fact, both signal and background have  the same  $\met$ distribution, see figure \ref{fig:hp_aa_nocut} (left panel). However, another variable useful to reduce the background  and enhance the signal is the transverse mass of the di-photon system, $M_T(\gamma\gamma)$  of eq.~(\ref{eq:MT}), see figure \ref{fig:hp_aa_nocut} (right panel): by vetoing the region with $M_T(\gamma\gamma) > 250$~GeV we can decreases the non-resonant background contribution significantly, as shown in table \ref{table:hp_aa}
(where the $h$ mass reconstruction is enforced as well). For this topology, all three signals considered are viable at the
standard LHC although kinematically they appear rather similar so that it may not be possible to distinguish one from the others. Further, as seen in figure \ref{fig:hp1_aa_nocut}, these are extracted by bulk regions of BLSSM parameter space
allowed by all available experimental constraints, notably by chargino searches at the LHC, which
require $m_{{\tilde{\chi}}_1^\pm}>250$ GeV by ATLAS \cite{Aad:2015jqa} and $m_{{\tilde{\chi}}_1^\pm}>210$ GeV by CMS \cite{Khachatryan:2014mma}. Hence, they are not particularly fine-tuned, rather they represent a genuine discovery scope afforded by this SUSY scenario over a substantial LSP mass range. 

Before closing this section, we should dwell shortly on the backgrounds we eventually considered. Clearly, one should certainly include the irreducible background from the associated production of the $Z$ boson and the SM-like Higgs state where the $Z$ decays to two neutrinos, which  resonates at $m_{\gamma\gamma}\approx m_h$.  There are also two other similarly resonant backgrounds. The first one is direct SM-like Higgs production and decay, but this is reducible since it does not have real  $\met$ (rather a mis-measured one from detector effects). The second one is the SM-like Higgs boson production in association with a $W^\pm$ state  where the latter decays to  lepton and neutrino (wherein the lepton is missed in the detector, again leading to additional mis-measured $\met$ alongside the one emerging from the neutrino). Moreover, a continuum background which also plays a role is $Z\gamma\gamma$, which in fact competes
with $Zh$. Finally, there are several non-resonant background sources that can mimic the signal when they have mis-measured $\met$ and happen to reconstruct two photons with an invariant  mass close to the mass of the SM-like Higgs boson, but they were found to be negligible: these were from QCD, $t\bar{t}$ and Drell-Yan production of two electrons. 

\begin{table}[t]
\begin{center}
{\small\fontsize{9}{9}\selectfont{
\begin{tabular}{|@{\hspace{0.02cm}}c@{\hspace{0.02cm}}||@{\hspace{0.02cm}}c@{\hspace{0.02cm}}|@{\hspace{0.02cm}}c@{\hspace{0.02cm}}|@{\hspace{0.02cm}}c@{\hspace{0.02cm}}|@{\hspace{0.02cm}}c@{\hspace{0.02cm}}|@{\hspace{0.02cm}}c@{\hspace{0.02cm}}|@{\hspace{0.02cm}}c@{\hspace{0.02cm}}|}
\hline
\multicolumn{2}{|@{\hspace{0.02cm}}c@{\hspace{0.02cm}}|}{}&\multicolumn{4}{c|}{Backgrounds}&Signal\\
\hline
\multicolumn{2}{|@{\hspace{0.02cm}}c@{\hspace{0.02cm}}|}{ Process}& $Z(\to\nu\bar{\nu})h$  & $W(\to l\bar{\nu})h$ &$h$& $Z(\to \nu\nu)\gamma\gamma$&$Z'\to Z(\to \nu\nu)h$ \\
\hline
\hline
\multicolumn{2}{|@{\hspace{0.02cm}}c@{\hspace{0.02cm}}|}{Before cuts} &37.9 & 66.4 & 6129 & 9126 & 0.269 \\
\hline
\multirow{4}{*}{\rotatebox{90}{Cut}}&$n(\gamma)\geq 2$ with $p_T(\gamma)> 20$~GeV and $|\eta(\gamma)|<2.5$  &27.79& 48.35& 4423.6 &1979.7& 0.119 \\
\cline{2-7}
&$110$~GeV $< m_{\gamma\gamma}<130$~GeV & 25.81 & 44.65 & 4298.7 & 152.9 & 0.0655  \\
\cline{2-7}
&veto on $l$ with $p_T(l)> 20$~GeV and $|\eta(l)|<2.5$ &25.80& 11.97& 4296.1& 152.8 & 0.0655  \\
\cline{2-7}
&$\slashed E_T>550$ GeV &0.0127& 0.00059& 0& 0.0192& 0.0459  \\
\hline
\end{tabular}}}
\end{center}
\caption{The cut flow on signal and background events for the $\gamma\gamma +\met$ signature in the $Z'$ mediator case. These events are generated at $\sqrt s=14$~TeV with ${\cal L}dt= 300$~fb$^{-1}$.}
\label{table:zpaa} 
\end{table}

\begin{table}[h!]
\begin{center}
{\small\fontsize{9}{9}\selectfont{
\begin{tabular}{|c||c|c|c|c|c|c|c|c|}
\hline
\multicolumn{2}{|c|}{}&\multicolumn{4}{c|}{Backgrounds}&\multicolumn{3}{c|}{Signal}\\
\hline
\multicolumn{2}{|c|}{ Process}& $Z(\to\nu\bar{\nu})h$  & $W(\to l\bar{\nu})h$ &$h$& $Z(\to\nu\nu)\gamma\gamma$&\multicolumn{3}{c|}{$h' \to h(\to\met) h$} \\
\hline
\hline
\multicolumn{2}{|c|}{Before cuts} &57.0 & 66.3 & 7200 & 8400 &\color{red}{386}&\color{blue}{347}&\color{green}{183} \\
\hline
\multirow{3}{*}{\rotatebox{90}{Cut}}&$\met > 100$~GeV &19.3& 8.51& 0.114 &547.2&\color{red}{73.1}&\color{blue}{62.0}&\color{green}{27.6} \\
\cline{2-9}
&$115$ GeV $ < m_{\gamma\gamma}< 130$ GeV & 13.0 & 5.32 & 0.065 & 28.7 &\color{red}{45.6}&\color{blue}{37.4}&\color{green}{14.6}  \\
\cline{2-9}
&$5$ GeV $ < M_{T}({\gamma\gamma}) < 250$ GeV &0.88& 0.51& 0.030& 2.01 &\color{red}{45.4}&\color{blue}{37.3}&\color{green}{14.5}  \\
\hline
\end{tabular}}}
\end{center}
\caption{The cut flow on signal and background events for the $\gamma\gamma +\met$ signature in the $h'$ mediator case. These events are generated at $\sqrt s=14$~TeV with ${\cal L}dt= 300$~fb$^{-1}$. In the signal column, the red (left) entries are for $m_{\tilde{\chi}_1^0}\simeq 11$~GeV, the blue (middle) entries are for $m_{\tilde{\chi}_1^0}\simeq 30$~GeV while the green (right) entries are for $m_{\tilde{\chi}_1^0}\simeq 49$~GeV.   }
\label{table:hp_aa} 
\end{table}
\subsection{The $4l$ $+$ $\met$ signature}

The $h\to ZZ^*\to 4l$ decay  ($l=e,\mu$) has a rather small rate but it also offers a very suppressed background and for this has always been considered as the golden channel for a Higgs boson discovery. Hence, no surprise it turns out to play a significant role also in mono-$h$ searches in the BLSSM. Let us illustrate  our selection strategy this time starting  from
the background channels one has to deal with, which  are as follows.
\begin{enumerate}
\item $Z(\to \nu\bar{\nu})h(\to ZZ^*)$, which is an irreducible background.
\item $Z(\to l\bar{l})h(\to ZZ^*)$, which is also an irreducible background (the secondary $Z$ is assumed to decay into neutrinos) and with a larger cross section than the previous one. 
\item $W(\to l\bar{\nu})h(\to ZZ^*)$, where the lepton from the $W^\pm$ (or indeed one of the others) is missed.
\item $h(\to ZZ^*)$ with $\met$ coming  from mis-measurements of soft radiation.
\end{enumerate}
Other backgrounds can come from three gauge boson production: i.e., $Z\gamma\gamma,~WWW,~ZWW,~ZZ\gamma$, but these are highly suppressed and can be neglected. (Also $t\bar t$
production and decay is negligible here.)

\begin{figure}[h!]
\begin{center}
\includegraphics[width=0.52\textwidth]{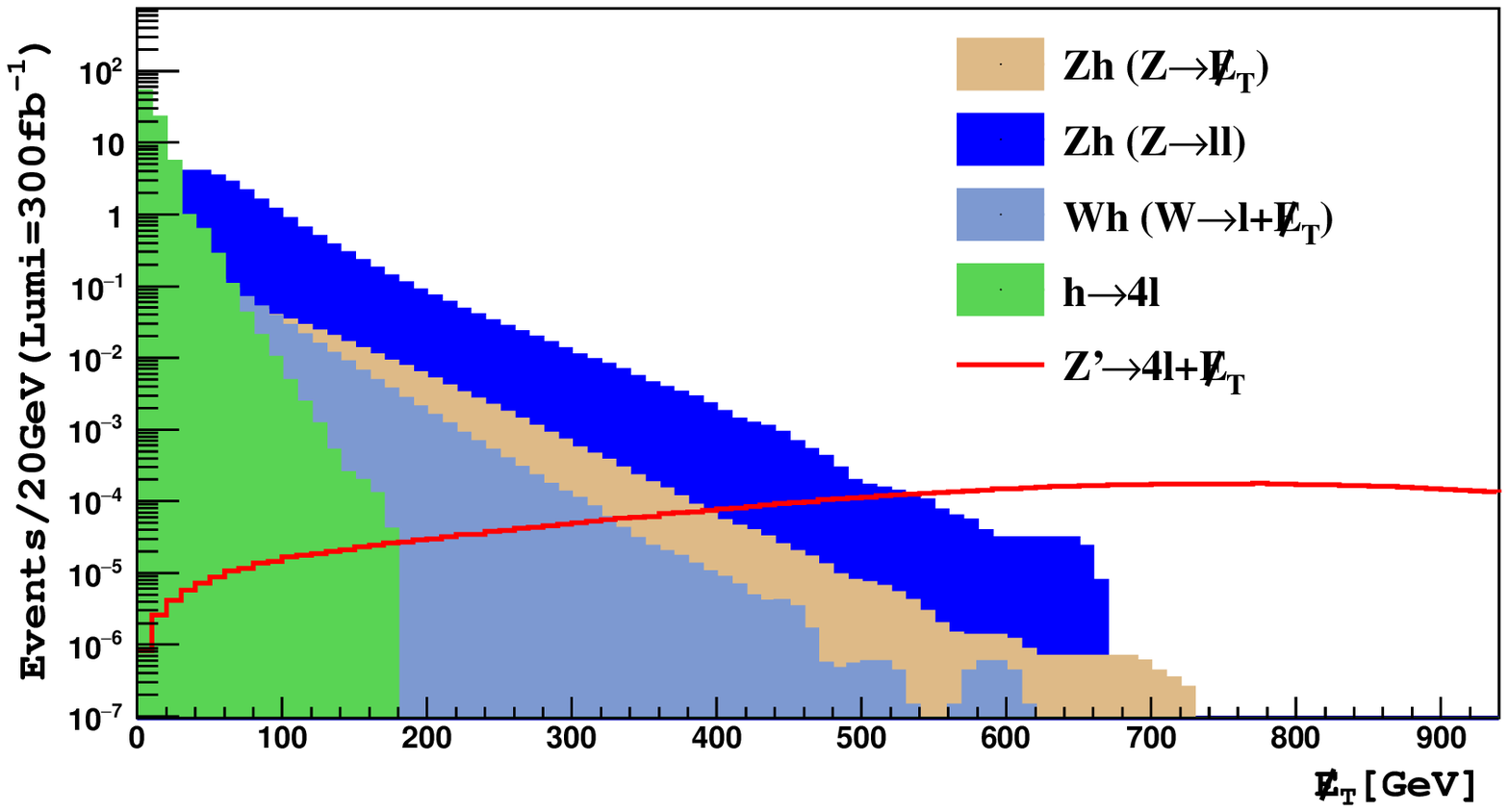}~~\includegraphics[width=0.52\textwidth]{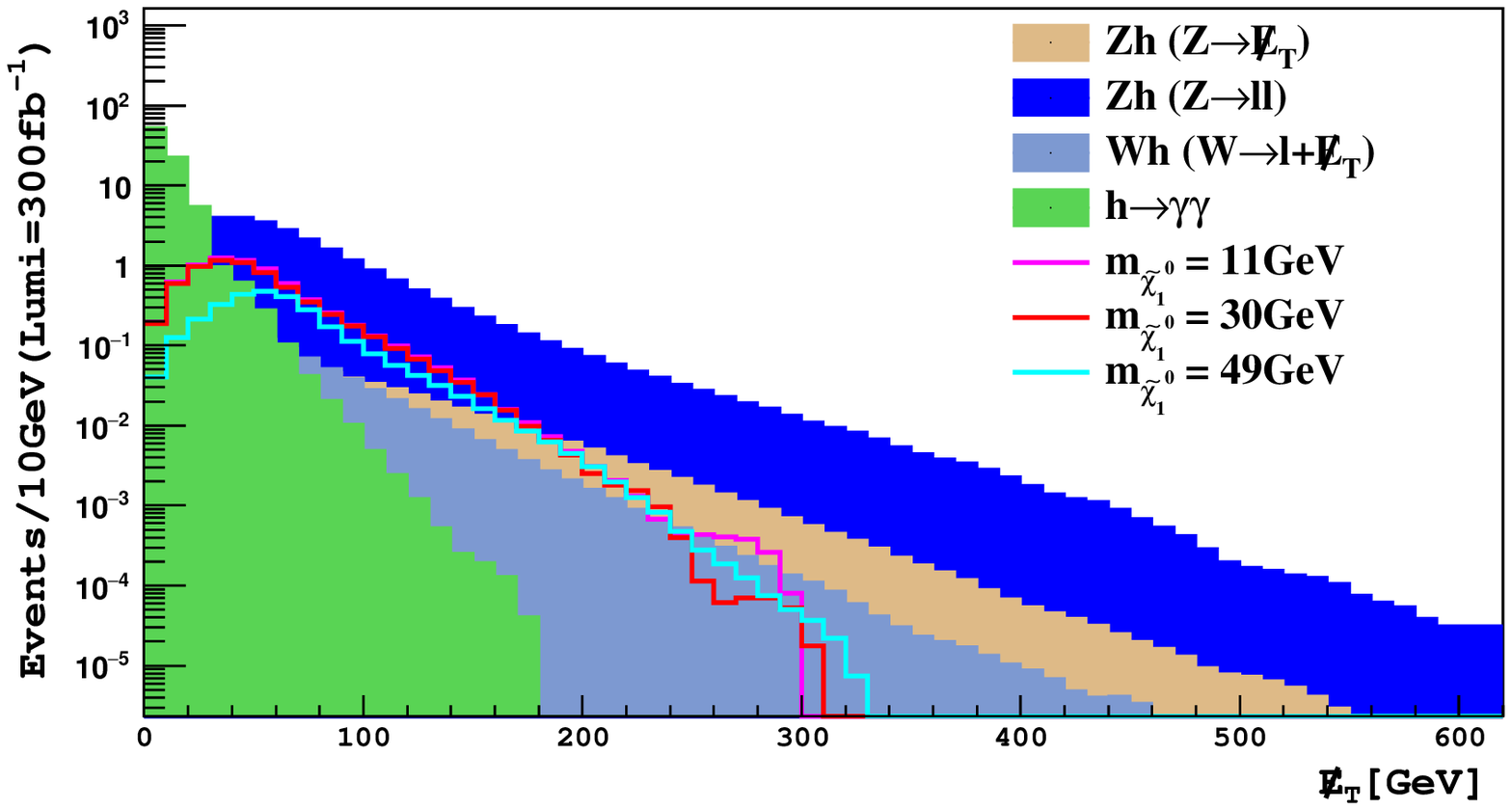}
\caption{Number of events of both signals ($pp\to Z'\to Zh\to 4l+$ $\met$ on the left plus $pp\to h^\prime \to hh\to 4l +\met$ in the right) and their relevant backgrounds generated at 
14 TeV and normalised per bin width after $300$~fb$^{-1}$ of integrated luminosity versus $\met$. The cuts described in the text have been applied here.} 
\label{fig:4l_nocut}
\end{center}
\end{figure}

Events are first required to have at least four reconstructed leptons, for which we used a selection based on the CMS four lepton discovery channel of the SM-like Higgs boson. Electrons are required to have a minimum $p_T$ of $7$~GeV and need to be in the pseudorapidity range $|\eta| < 2.5$ (which is the geometrical acceptance of both the ATLAS and CMS experiments, roughly). Selected muons need to be reconstructed with $p_T > 6$~GeV and also be in the same geometrical acceptance. All four leptons have to be isolated. The isolation variable is defined as the sum of the transverse momenta of the tracks inside a cone of opening $\Delta R \ge 0.3$ around the lepton. This variable is known to be robust against an  increase in the number of pileup interactions and mis-identification issues. The cut on the isolation variable was optimised by using the lowest $p_T$ lepton for each signal. Leptons of opposite sign and same flavour are then paired and any such di-lepton system is required to have an invariant mass larger than $4$~GeV in order to suppress the light-jet QCD background.  If more than two di-lepton pairs can be formed, ambiguities are resolved as follows: the di-lepton system with total invariant mass closest to the $Z$ boson mass is chosen as the first $Z$ boson. Among all valid, i.e., same flavour opposite sign, di-leptons that can be formed from the remaining tracks, we choose as the second $Z$ boson the di-lepton system with the highest $p_T$ whose total three-momentum  vector is at least $\Delta R \ge 0.05$ away from the first di-lepton. This set of selections is applied to both channels, i.e.,  $Z'$  and $h'$ topologies, while the difference between the two signals can be exalted by choosing additional kinematic cuts, different from one case to the other. 

The possible choice is in principle guided by figure \ref{fig:4l_nocut}, which is constructed after
the above cuts are enforced on both $Z'$ and $h'$ topologies, In practise, though, by looking at the plot on the right-hand side,
it is clear that the $Z'$ mediator case is again irrelevant numerically, so that we will not treat it any further here. For the $h'$ mediator case, the transverse mass variable is again highly effective to reduce the noise, thus we select  events with transverse mass in range $[115,~250]$~GeV and invariant mass of the four reconstructed leptons in the range $[115,~130]$~GeV. At the same time we reject an event if it has missing transverse energy less than $20$~GeV, as shown in table \ref{table:hp_4l}, wherein the $h$ mass is also reconstructed from the four leptons). Even if less than in the
case of the di-photon channel, also the four-lepton rate from mono-$h$ in the BLSSM is significant at the LHC in standard
running condition, so it can be used to supplement a potential discovery herein.

\begin{table}[h!]
\begin{center}
{\small\fontsize{9}{9}\selectfont{
\begin{tabular}{|c||c|c|c|c|c|c|c|c|}
\hline
\multicolumn{2}{|c|}{}&\multicolumn{4}{c|}{Backgrounds}&\multicolumn{3}{c|}{Signal}\\
\hline
\multicolumn{2}{|c|}{ Process}& $Z(\to \nu\bar{\nu})h$  & $Z(\to l\bar{l})h$ &$W(\to l\bar{\nu})h$& $h$&\multicolumn{3}{c|}{$h' \to h(\to\met) h$} \\
\hline
\hline
\multicolumn{2}{|c|}{Before cuts} &0.654 & 30.0 & 1.11 & 112 & \color{red}{7.46}&\color{blue}{7.03}&\color{green}{2.95} \\
\hline
\multirow{3}{*}{\rotatebox{90}{Cut}}&$\met > 20$~GeV &0.613&26.84&0.965&7.56 & \color{red}{6.68}&\color{blue}{6.27}&\color{green}{2.79} \\
\cline{2-9}
&$115$ GeV $ < m_{4l}< 130$ GeV &0.282&0.241&0.308&0.82 & \color{red}{1.62}&\color{blue}{1.53}&\color{green}{0.681}  \\
\cline{2-9}
&$115$ GeV $< M_T(4l)< 250$ GeV &0.177&0.207&0.199&0.82 & \color{red}{1.62}&\color{blue}{1.53}&\color{green}{0.681}  \\
\hline
\end{tabular}}}
\end{center}
\caption{The cut flow on signal and background events for the $4l+\met$ signature in the $h'$ mediator case. These events are generated at $\sqrt s=14$~TeV with ${\cal L}dt= 300$~fb$^{-1}$. In the signal column, the red (left) entries are for $m_{\tilde{\chi}_1^0}\simeq 11$~GeV, the blue (middle) entries are for $m_{\tilde{\chi}_1^0}\simeq 30$~GeV while the green (right) entries are for $m_{\tilde{\chi}_1^0}\simeq 49$~GeV.}
\label{table:hp_4l} 
\end{table}

\section{Conclusions}
\label{sec:con}

We have considered the scope of current mono-$h$ searches in probing  a non-minimal SUSY scenario, the BLSSM, which offers key advantages with 
respect to the MSSM in relation to its ability to naturally embed massive neutrinos. Rather than concentrating on mono-$h$
topologies which are common with the MSSM though, we have instead focused on those which are specific to the BLSSM. As the latter, in comparison to the former, possesses (amongst other states) an
 additional heavy neutral gauge boson ($Z'$, with mass of ${\cal O}(2~{\rm TeV})$) 
as well as an intermediate Higgs  ($h'$, with mass of ${\cal O}(0.2~{\rm TeV})$) states, both of which may be within the LHC reach,
we looked in particular at the topologies onsetting the two production and decay channels $pp\to Z'\to Zh\to 4l+$ $\met$ and $pp\to h^\prime \to hh\to 4l +\met$, which indeed see a $Z'$ and $h'$ as mediators, respectively, of DM pair production (alongside that of neutrinos), this being the lightest neutralino. We have therefore tested the scope of the two most precise
decay of the $h$ state, into di-photons and $Z$-boson pairs, in extracting excesses attributable to the BLSSM above and
beyond the yield of the SM. After a refined MC analysis based on multi-parton scattering, parton shower, hadronisation/fragmentation as well as detector effects, we have been able to show that a significant excess can be established by the end of the LHC Run 2
in both $h$ decay channels in the case of the $h'$ mediated topology, but not for the case of the $Z'$ mediated one.
A key to achieve this has been the fact that the heavier $h'$ masses involved with respect to the one of the $Z$ 
boson (the mediator of mono-$h$ events in the MSSM) afford one with rather selective criteria in improving the $S/B$
ratio.
This phenomenology occurs only for rather light LSP masses, in the range up to $m_h/2$ (as the relevant topology
proceeds via a $h$ decay into DM pairs), yet all still allowed experimentally. Further, despite the significance of all benchmarks tested, it is not possible to extract (neither in terms of total event rates nor in terms of kinematic differences) 
the mass of the DM candidate. Nonetheless, our results can inform experimental searches aimed at extracting mono-$h$
signals of DM with a potential non-minimal SUSY nature in the foreseenable future, or else in imposing strong bounds 
on their existence.   
\section*{Acknowledgments}
The work of W.A. and S.K. is partially supported by the STDF project 18448, the ICTP Grant AC-80 and the European Unions Horizon 2020 research and innovation programme under the Marie Sklodowska-Curie grant agreement No.~690575. A.H. is partially
supported from the STDF project 6109 and the EENP2 FP7-PEOPLE-2012-IRSES grant. S.M. is financed in part through the NExT Institute. All authors are supported by the grant H2020-MSCA-RISE-2014 No.~645722 (NonMinimalHiggs).

\end{document}